# *d*-zero Magnetism in Nanoporous Amorphous Alumina Membranes

*Amir Sajad Esmaeily, M. Venkatesan, S. Sen and J. M. D. Coey\**

School of Physics and CRANN, Trinity College, Dublin 2, Ireland.

**Abstract:** Nanoporous alumina membranes produced by mild or hard anodisation in oxalic acid at potentials ranging from 5 – 140 V have a controllable pore surface area of up to 200 times the membrane area. They exhibit a saturating magnetic response that is temperature-independent and almost anhysteretic below room temperature. The magnetism cannot be explained by the ~1 ppm of transition-metal impurities present in the membranes. The magnetic moment increases with the area of the open nanopores, reaching values of 0.6 Bohr magnetons per square nanometer for mild anodisation and 8 $\mu_B \text{nm}^{-2}$ for hard anodisation where the growth rate is faster. Crystallization of the membrane or chemical treatment with salicylic acid reduces the effect by up to 90 %. The magnetism is therefore associated with the open pore surfaces of the amorphous $Al_2O_3$, and it is independent of the orientation of the applied field with respect to the membrane surface. Possible explanations in terms of electrons associated with oxygen vacancies (F or $F^+$ centres) are discussed and it is concluded that the effect is due to giant orbital paramagnetism rather than collective ferromagnetic spin order.

Keywords: *d*-zero magnetism, amorphous alumina, porous anodized alumina, giant orbital paramagnetism.

Correspondence: jcoey@tcd.ie

Popular Summary

**_d_-zero magnetism in nanoporous amorphous alumina membranes.**

Magnetic materials have been with us for thousands of years, and we have known for almost 90 years that their magnetism, at least at room temperature, can be traced to the 'spin' of certain electrons – the 3$d$ electrons – that behave like tiny elementary magnets. Iron and cobalt, for example, are rich in 3$d$ electrons and they couple together to form a collective ferromagnetic state.

Here we have discovered a material that looks like a ferromagnet, yet it possesses no 3d electrons. The material in a non-crystalline variety of sapphire, in the form of a thin membrane riddled with an array of tiny holes. These pores increase the surface area about 200 fold, up to about a square meter per gram; we show that a new type of '$d$-zero' magnetism arises from the huge pore surface.

Our explanation is that the electrons at the pore surfaces are locked together in a coherent quantum state where they orbit together in the same direction in response to a magnetic field. The reason is traced to the zero-point energy of the vacuum. Empty space is seething with energy, but normally it has almost no observable effects. $d$-zero magnetism is an exception.

Our work may lead condensed-matter physicists to look for other manifestation of this mysterious and all-pervasive background radiation, a cosmic dance that fills space and in which all life has evolved.

# 1. Introduction

$d$-zero magnetism is the ferromagnetic-like response to an applied magnetic field of a material that is formally devoid of the unpaired $3d$ electrons usually needed for high temperature magnetic order [1, 2]. The effect is weak — a typical order or magnitude at room temperature is a Bohr magneton per square nanometer of surface, with little or no temperature dependence. There is no generally-accepted explanation. Doubts persist about the reality of the phenomenon because there are a number of well-documented artefacts and impurity effects [3-9] that might explain some of the measurements.

Many of the materials that display the effect are oxides. The effect is generally absent or undetectable in well-crystallized bulk material. It appears in polycrystalline granular solids [10], fine powders [11], thin films [12, 13], single-crystal surfaces [11, 14] and nanoparticles [15-19]. $CeO_2$ is the example that has been more intensively studied [20]. Much evidence suggests that $d$-zero magnetism is a surface or interface phenomenon, which is why the magnitude of the saturation moment is often related to the surface area. There are about 18 surface oxygen ions per nm$^2$ at a close-packed oxygen surface, but the surfaces, especially in ambient conditions, are difficult to characterize because they tend to reconstruct and attract adlayers of water. Experience with oxide interfaces, which are liable to exhibit emergent properties such as metallic conduction or magnetism that is absent from either of the constituents, alerts us to the possibility of a different electronic state at the surface. The $LaAlO_3/SrTiO_3$ interface is a famous example [21-23]; it can be metallic or magnetic yet neither oxide by itself has these properties in the bulk (although irradiated $SrTiO_3$ crystals [24] and the surface of crystalline $SrTiO_3$ [11] have recently been shown to exhibit features typical of $d$-zero magnetism). Some form of reconstruction, whether by charge transfer or atomic rearrangement, often in the form of oxygen vacancies, is inevitable at an extended polar surface or interface in order to eliminate a divergence in the energy known as the polar catastrophe that is associated with the electric field created by the distribution of surface charge.

Nanoporous oxide membranes are very interesting in this respect. The ratio $\rho$ of internal surface area of the pores to the area of the two flat surfaces of a membrane of thickness $t_m$ with a regular hexagonal array of pores of radius $r$ and spacing $d$ is

$$\rho = 2\pi r t_m/\sqrt{3}d^2 \qquad (1)$$

For example, in a membrane with $d$ = 140 nm, $r$ = 30 nm and $t_m$ = 40 μm, this ratio exceeds 200. Such membranes thicker than 90 nm have most of their surface area in the pores.

Micron-thick membranes therefore provide an excellent opportunity to enhance the surface/volume ratio in order to study surface magnetism in oxides. To achieve the same enhancement by reducing a crystal of the same thickness $t_m$ to powder particles of radius $r_p$ requires

$$\rho = 3t_m/2r_p \qquad (2)$$

or $r_p \approx 290$ nm for $t_m = 40$ μm. It is relatively straightforward to produce a nanoporous oxide membrane electrochemically, but difficult to obtain a 600 nm ceramic powder by milling, although there are options to synthesise nanoparticles directly.

There are a few reports in the literature on the magnetism of nanoporous $Al_2O_3$ [25], $TiO_2$ [26] and $Cu_2O$ [27] and $Cu_2O$/porous anodic alumina [28] films. Most remarkable is the report of H. Y. Sun and co-workers [25], who produced nanoporous alumina membranes of submicron thickness by brief mild anodisation at 45 V in oxalic acid. These membranes exhibited ferromagnetic-like signals in SQUID magnetometry that were as high as 60 kAm$^{-1}$ when the magnetic field was applied perpendicular to the plane of the membrane, but there was a linear, paramagnetic response when the field was applied in-plane (For comparison, the magnetization of Ni is 500 kAm$^{-1}$).

Here we have conducted a systematic investigation of the magnetic properties of nanoporous amorphous alumina membranes prepared by a mild [29] or hard [30] anodization process. We find magnetic signals that are more than an order of magnitude or more greater than anything that could be explained by the 3$d$ impurities that are present at a level of less than 1 ppm in our samples. We associate the magnetism with internal pore surfaces, and discuss possible explanations of its origin and sign.

## 2. Experimental Methods and Materials

The aluminium foil used for anodisation was supplied by Advent Research Materials (UK). It is 250 μm thick and nominally 5N (99.999%) pure with stated impurity levels of ferromagnetic elements Fe < 0.7 ppm and Ni < 0.6 ppm. 12 mm discs were cut with a steel punch, and the discs were mounted in a Teflon cell where the central 7 mm of the disc was exposed to the electrolyte, forming the anode in an electrolytic cell.

Four sets of samples were prepared by the mild method in 0.3 M oxalic acid at 17°C at voltages ranging from 5 V – 70 V. A first step of anodisation was carried out at the selected voltage for five hours. The disordered oxide layer so formed was then removed using a solution containing 0.2 M chromic acid and 0.5 M phosphoric acid at 60°C for 3 hours. The dimpled aluminium was then re-anodized in a second-step for a further three hours at the selected voltage.

The four sets were continued into the hard anodisation regime, in a single-step process, starting at 40 V for 10 min, and then ramping up the potential at 30 V/min to a selected value ranging from 80 V – 140 V, where it held for 20 min. Argon was bubbled through the electrolyte during anodisation of the third set, and extra voltage points were included in the fourth set. Anodisation rates are an order of magnitude faster in the hard regime.

Another two sets of thinner, mild-anodized samples was prepared at 40 V with same first step but varying the second-step anodisation time from 2 to 60 minutes to obtain two series of four very thin samples of different thicknesses prepared in the same conditions.

Small specimens (1 mm$^2$) used for analysis in a Zeiss Ultra scanning electron microscope (SEM) were cut with Ti scissors and coated with 2-5 nm of gold . Membrane thickness was determined by direct SEM observation of a fractured surface at a kink in the membrane.

X-ray diffraction was performed on larger 20 mm discs, which were prepared in another cell. Photoluminescence was measured on one complete series of 7 mm diameter membranes using 405 nm excimer laser excitation and on several larger discs. Selected samples, anodized at 20 V, 70 V and 140 V were measured by electron paramagnetic resonance (EPR).

The specimens used for magnetic analysis were squares roughly 5 × 5 mm$^2$ cut from the centres of the discs with titanium scissors. When following this protocol, these samples were shown to be free of contamination by the iron that is introduced around the rim of the disc from the steel punch. The samples were mounted in straws for measurement in a 5 T Quantum Design SQUID magnetometer. After cutting, the samples were handled with tools made of wood or plastic.

All samples in the second set were analysed for traces of Fe, Co and Ni using Laser Ablated Inductively-Coupled Mass Spectroscopy (LA-ICP-MS).

## 3 Experimental Results

Images of the membranes grown in the mild and hard regimes at different voltages are presented in Fig. 1. Pores increase in size with voltage, and their positions are irregular for voltages below 30 V. The arrays becomes periodic at 30 – 50 V and are most ordered at 40 V, but large ∼ 100 nm irregular pores appear in the mild regime above 50 V. Pores in the hard regime are about 50 nm in diameter, and more widely spaced with less dependence on the anodic voltage. Fig. 2 shows plots the pore diameter $2r$, average pore spacing $d$, and membrane thickness $t_m$ as a function of voltage, covering both regimes. Unlike the mild case, the underlying pore structure was not visible from at the top surface of the hard anodized samples, so samples had to be imaged from the bottom side. The aluminium substrate was chemically etched away using a solution comprised of equal amounts of saturated $CuSO_4$ and HCl, and the alumina barrier layer was then removed by argon ion milling to expose the pores.

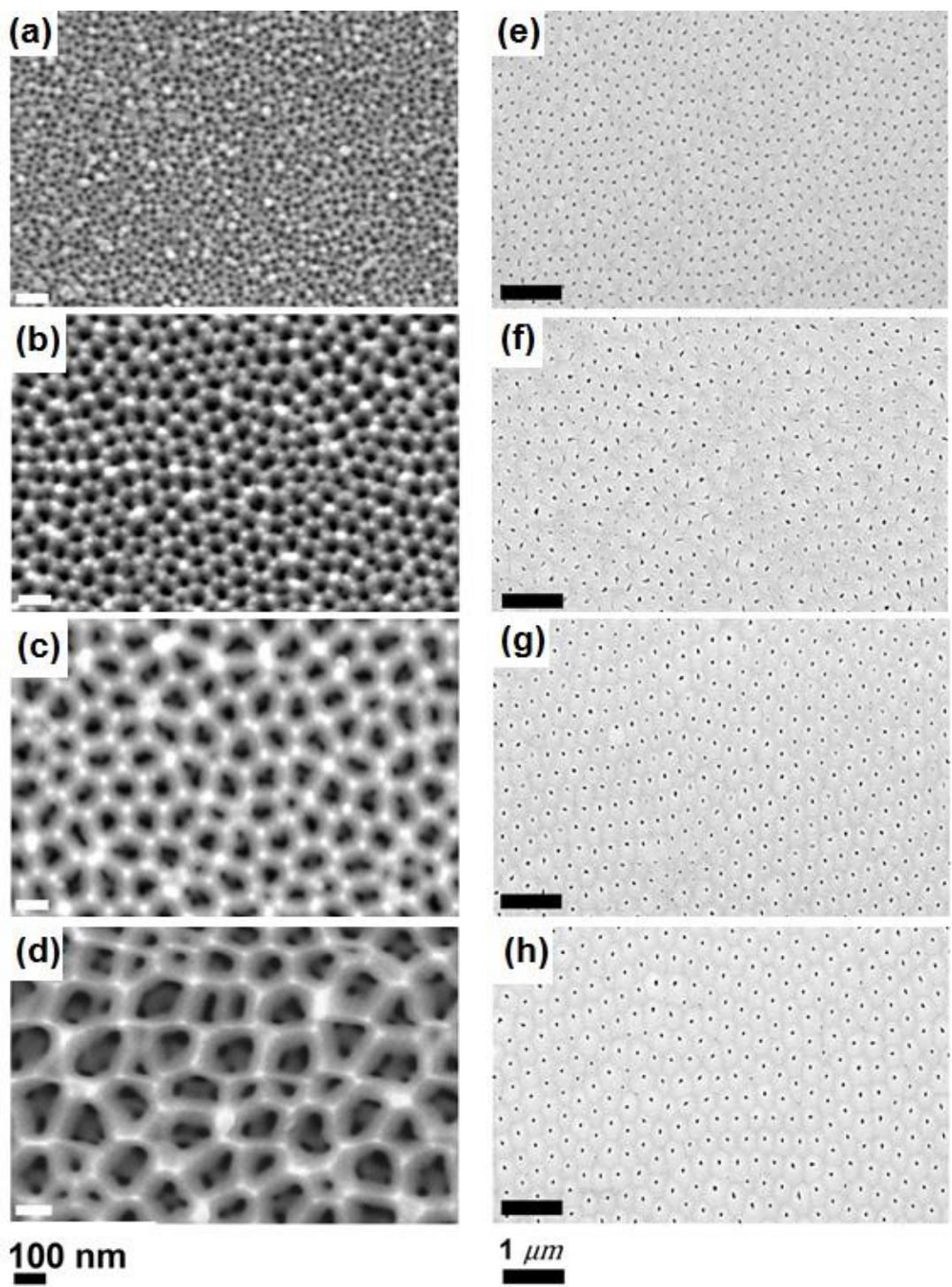

Figure 1. Evolution of pore structure with anodisation voltage. The left column is for mild anodisation at 17 °C (10, 30, 50, 70 V, scale bars 100 nm), the right column is for hard anodisation at 0 °C (80, 100, 120, 140 V, scale bars 1 micron). Details of sample preparation are given in the text.

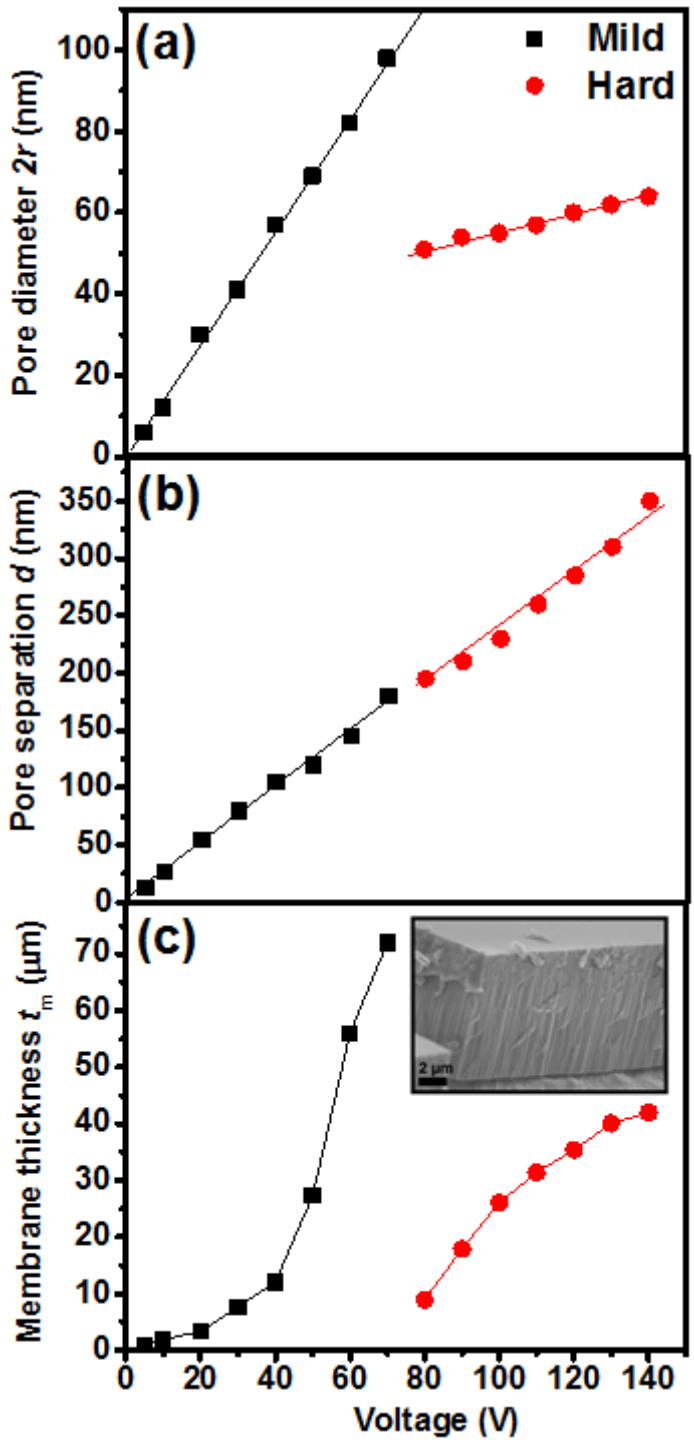

Figure 2. Plots of a) average pore diameter, b) average inter-pore distance and c) membrane thickness as a function of voltage, spanning the mild and hard anodisation regimes. The insert in c) shows the cross section of a typical membrane fracture surface (80 V) used for thickness measurements.

X-ray diffraction patterns of both types of membranes after removing the Al substrate are illustrated in Figure 3. They are amorphous, but crystallize in the cubic γ-Al$_2$O$_3$ structure when heated to 1000 °C in 10$^{-7}$ mbar. Density measurements on mild- and hard-anodized membranes yield values of 2054 and 2291 kgm$^{-3}$ (including the pores), respectively. Comparison with the value of 3300 kgm$^{-3}$ for random dense-packed amorphous alumina [31], indicates that the membranes have a free volume of approximately 25 %, in addition to their open porosity. The free volume contributes to the remarkable plasticity of this material, which aids the formation of the regular arrays of nanopores [32].

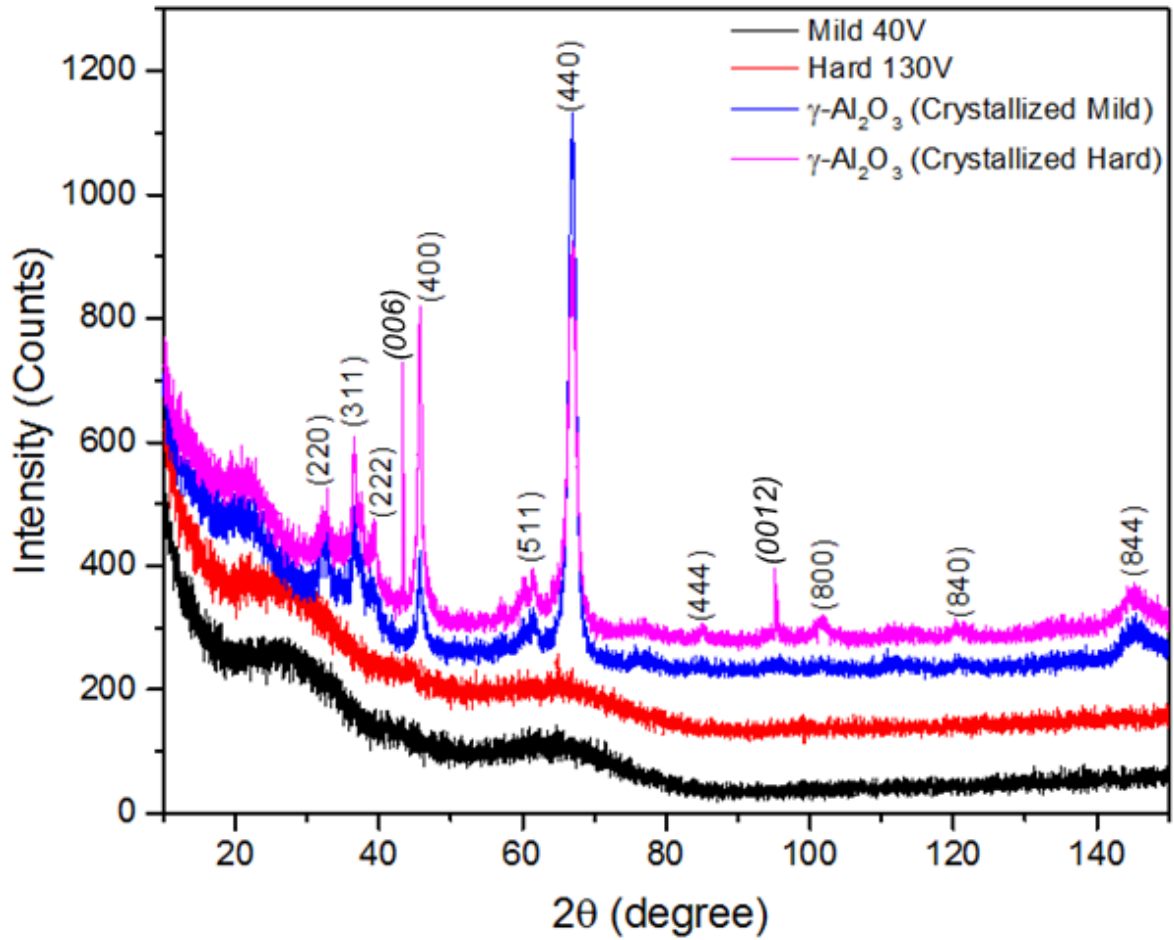

Figure 3. X-ray diffraction patterns of amorphous alumina membranes produced at 70 V and 130 V. The pattern after heating at 1000 °C in vacuum for 120 minutes shows crystallization of the membrane into γ-Al$_2$O$_3$. The four data sets are offset by 50 counts for clarity. α-Al$_2$O$_3$ peaks are indexed in italics.

Photoluminescence spectra (Figure 4) exhibit a peak at 520 nm, which tends to increase in intensity with membrane thickness in the mild and hard regimes, although the photoluminescence intensity is also found to increase monotonically with inter-pore spacing (Figure 4b).

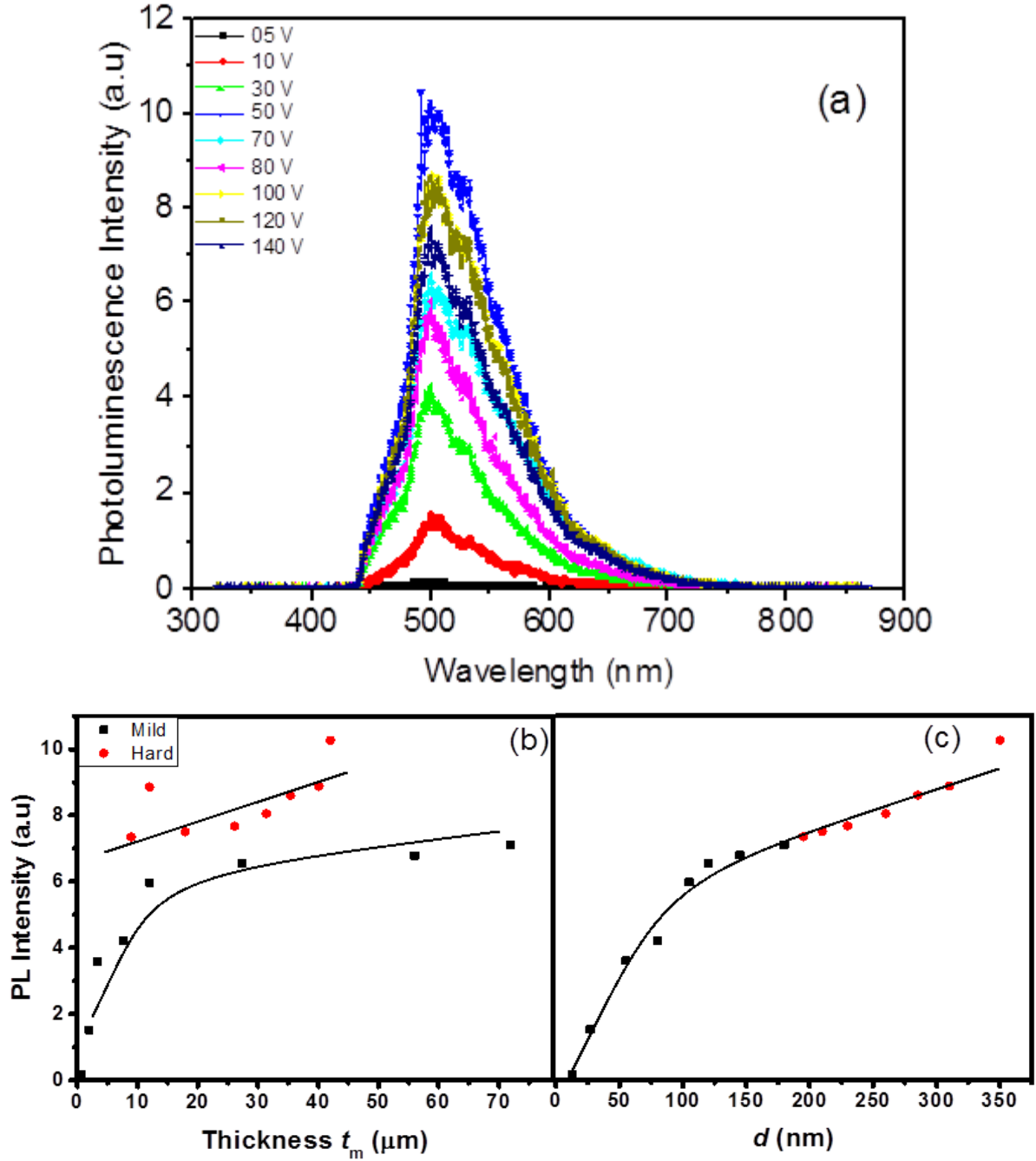

Figure 4. a) The photoluminescence spectra of the fourth set of amorphous alumina membranes prepared at different anodisation voltages with an excitation wavelength of 405 nm, and the line intensity plotted as a function of membrane thickness (b) and interpore spacing (c). The solid lines are guides to the eye.

The magnetization curve of the unoxidized aluminium illustrated in Fig 5a) shows its paramagnetic mass susceptibility of $\chi_m$(Al) = 7.3×10$^{-9}$ m$^3$kg$^{-1}$, in agreement with the literature value of 7.6×10$^{-9}$ m$^3$kg$^{-1}$, This susceptibility increases by 12% on cooling from 300 K to 4 K, but there is no sign of any paramagnetic Curie law upturn associated with isolated paramagnetic ions in the metal foil. Isolated Fe atoms in Al are nonmagnetic, with a Kondo temperature of order 5000 K [33] . The linear susceptibility of the foil is superposed on a very small saturating magnetic moment of about $1.5 \times 10^{-9}$ Am$^2$, which corresponds to 0.3 ppm of ferromagnetic iron impurity, an amount that is compatible with the nominal (< 0.7 ppm) and measured (0.5 ppm) iron content of the foil. All magnetic moments quoted here in Am$^2$ are normalized to a sample area of 25 mm$^2$.

Magnetization curves for two representative samples, one a thin membrane anodized at 10 V and the other a thick membrane anodized at 140 V are shown in Figure 5c) and d). The data are corrected for the paramagnetism of the aluminium substrate ($\chi$ = 19.6 10$^{-6}$), which gives a contribution of about $200 \times 10^{-9}$ Am$^2$ in 2 T, and the contribution of the alumina itself, which is diamagnetic ($\chi$ = -19.1 10$^{-6}$) but has a ferromagnetic moment, and a paramagnetic Curie-law contribution at low temperatures. a) and c) show the same slope, but the temperature-dependent scan of the moment of a membrane and substrate in fixed field in Fig 5b shows the upturn at low temperature, due to a Curie-law term superposed on an almost constant temperature-independent 'ferromagnetic-like' signal and the net paramagnetic background from aluminium and alumina. The Curie-law term would correspond to the presence of about 0.1 ppm of paramagnetic Fe$^{3+}$ in the alumina, or 1.2 ppm of free spins.

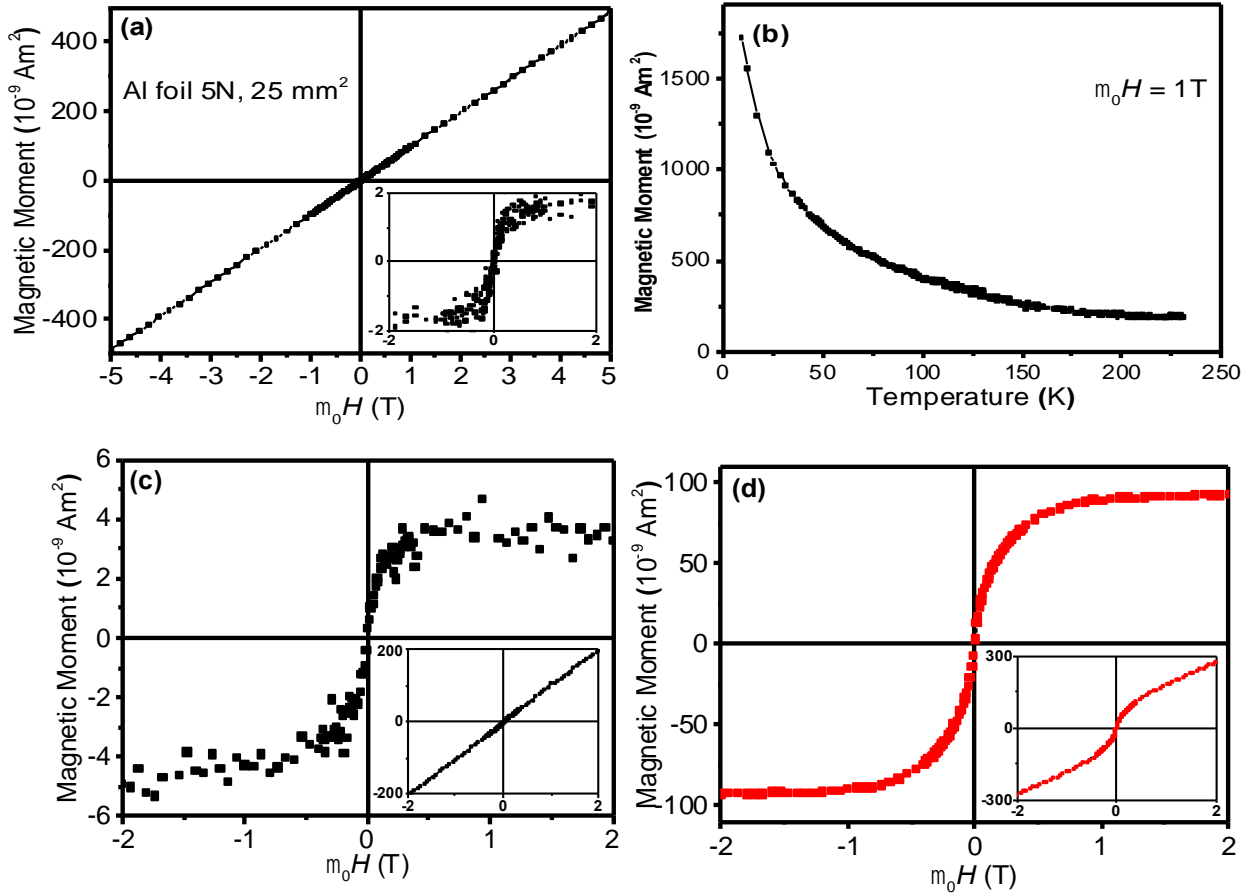

Figure 5. Magnetization curves for aluminium and for two alumina membranes. All moments shown are for 5 × 5 mm² samples. a) Moment of the aluminium foil on which the membranes are grown, (Insert) small residual signal after correction for the paramagnetic susceptibility, corresponding to 0.3 ppm of ferromagnetic impurity b) thermal scan of the moment in 1 T for an aluminium foil anodized at 70 V. Room-temperature magnetization curves for a mild- (10 V) (c) and hard-anodized membranes (140 V) (d) from set 4. The data are corrected for the paramagnetic slope; uncorrected data are shown in the inserts.

Generally the magnetic signals from the anodized membranes after correction for the susceptibility of the underlying unanodized aluminium exhibit a 'ferromagnetic-like' response to the applied magnetic field, which saturates in a field of approximately 0.5 T. The signal is 3 – 60 times greater than anything that could be due to ferromagnetic contamination of the aluminium itself.

Further examples for a mild-anodized and a hard-anodized membrane are shown in Fig 6, where the magnetization curves measured at 300 K and 4 K are compared. The observed coercivity is minimal, < 5 mT at both temperatures, and effectively zero within the error expected after saturating the magnetization in 5 T. The values measured after magnetizing in 1 T are < 5 mT. Just as remarkable is the absence of temperature dependence of the magnetization curve between 300 K and 4 K. Furthermore, the magnetization curve is practically identical for these membranes, regardless of whether the field is applied parallel or perpendicular to the membrane surface (Fig 6c)).

For the two extra series of thin ($t < \sim 1\ \mu m$) 40 V mild anodized membranes, the moments are proportionately smaller than for the thick membranes, and there was no anisotropy of the magnetization curves, according to whether the field is applied perpendicular or parallel to the surface.

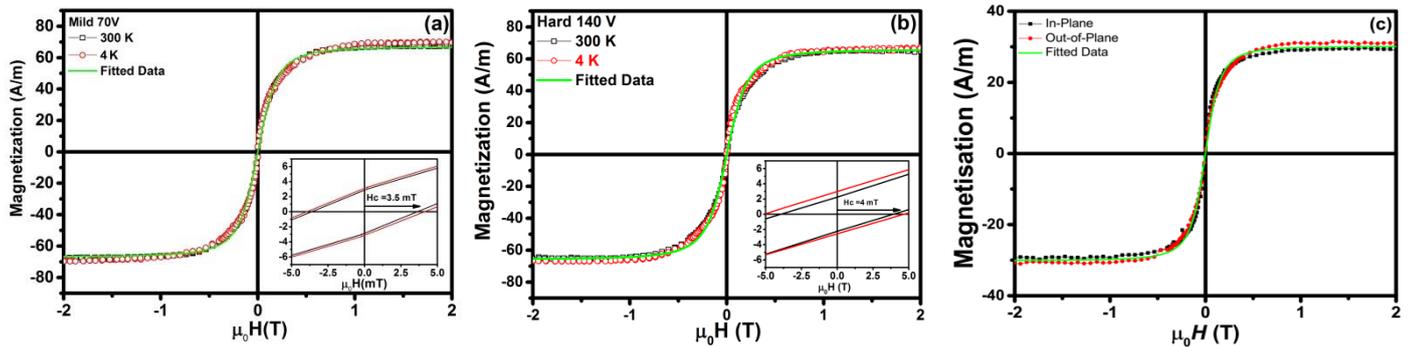

Figure 6. Comparison of magnetization curves at 300 K and 4 K for high-moment specimens anodized at (a) 70 and (b) 140 V. (c) Comparison of magnetization curves measured with the field, applied parallel and perpendicular to the surfaces of relatively thick membranes anodized at 50 V. The solid green curve is fit to the theoretical function $M = M_s\ x/\sqrt{(1 + x^2)}$, where $x=C\mu_0 H$.

A summary of magnetic moment data for the four sets of membranes, normalized to a common area of 25 mm², is plotted on a log scale as a function of anodisation voltage and porosity ratio in Figure 7. For both hard- and mild-anodized membranes there is a tendency for the moment to increase with anodisation voltage and membrane thickness. Although the values are rather scattered, the form of magnetization curve remains the same. In the plot of moment versus porosity in Fig 7b) the hard-anodised membranes are much more magnetic, reflecting their faster growth.

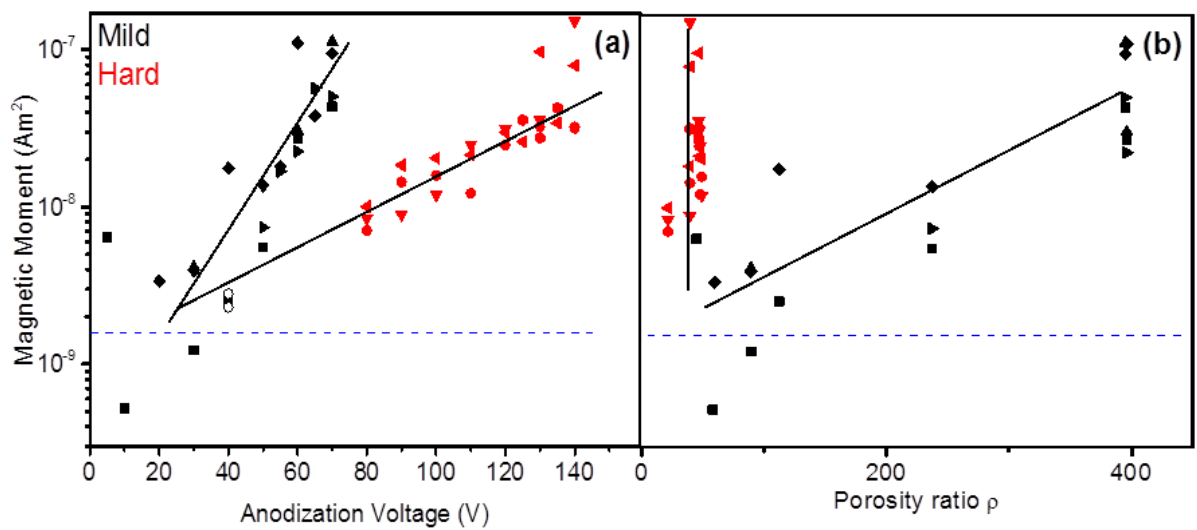

Figure 7. a) Compendium of magnetic moment data for the four sets of samples (solid symbols). b) a plot of magnetic moment on a logarithmic scale as a function of porosity ratio for mild (black) and hard (red) anodized samples. The horizontal dashed line marks the moment due to iron contamination of the aluminium. The open circles in a) are for membranes anodized at 40 V for an hour. Solid lines are guides to the eye.

The effect of crystallizing the amorphous alumina is illustrated in Fig. 8. Much of the porosity is eliminated, and the magnetic moment is largely destroyed by this treatment. A series of experiments that involved subjecting a 60 V membrane to a sequence of soaking in water or alcohol followed vacuum treatment, led to smaller reductions of magnetic moment, ranging from 10 – 80%. Most effective is immersion of the membrane in a 1 M aqueous solution of salicylic acid ($C_7H_6O_3$) for one hour, which has the effect of reducing the moment by 60 - 90 %. Salicylic acid is recognized as having a strong affinity to aluminium oxide surfaces [34]. All these experiments provide good evidence of a surface origin of the magnetism. Buried impurities or secondary magnetic oxide phases with a high Curie temperature would be insensitive to such treatment.

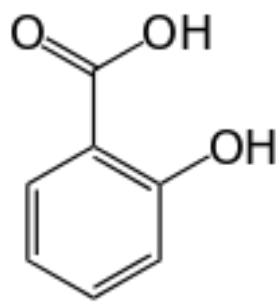

Salicylic acid

The LA-ICP-MS analysis of the 3$d$ impurities in one series membranes (36 data points) showed no trend with anodisation voltage. Averages, with standard deviations, were Fe 0.49(37) ppm, Co 0.01(2) ppm and Ni 0.05(6) ppm. Most of the measured moments are one or two orders of magnitude greater than could be accounted for by these ferromagnetic impurities, as we already demonstrated in Fig. 5 and Fig 7.

Finally, Electron Paramagnetic Resonance (EPR) spectra were measured for three of the membranes, anodized at 20 V, 70 V and 140 V and freed from their aluminium substrates. Narrow absorption lines with a linewidth of 1.5 mT and $g$ = 2.00396 are observed for the three specimens, with little change with orientation of applied field relative to the membrane. The numbers of free spins per gram for the three specimens were estimated from the peak areas as $2.5 \times 10^{16}$, $4.6 \times 10^{17}$ and $7.4 \times 10^{17}$ respectively.

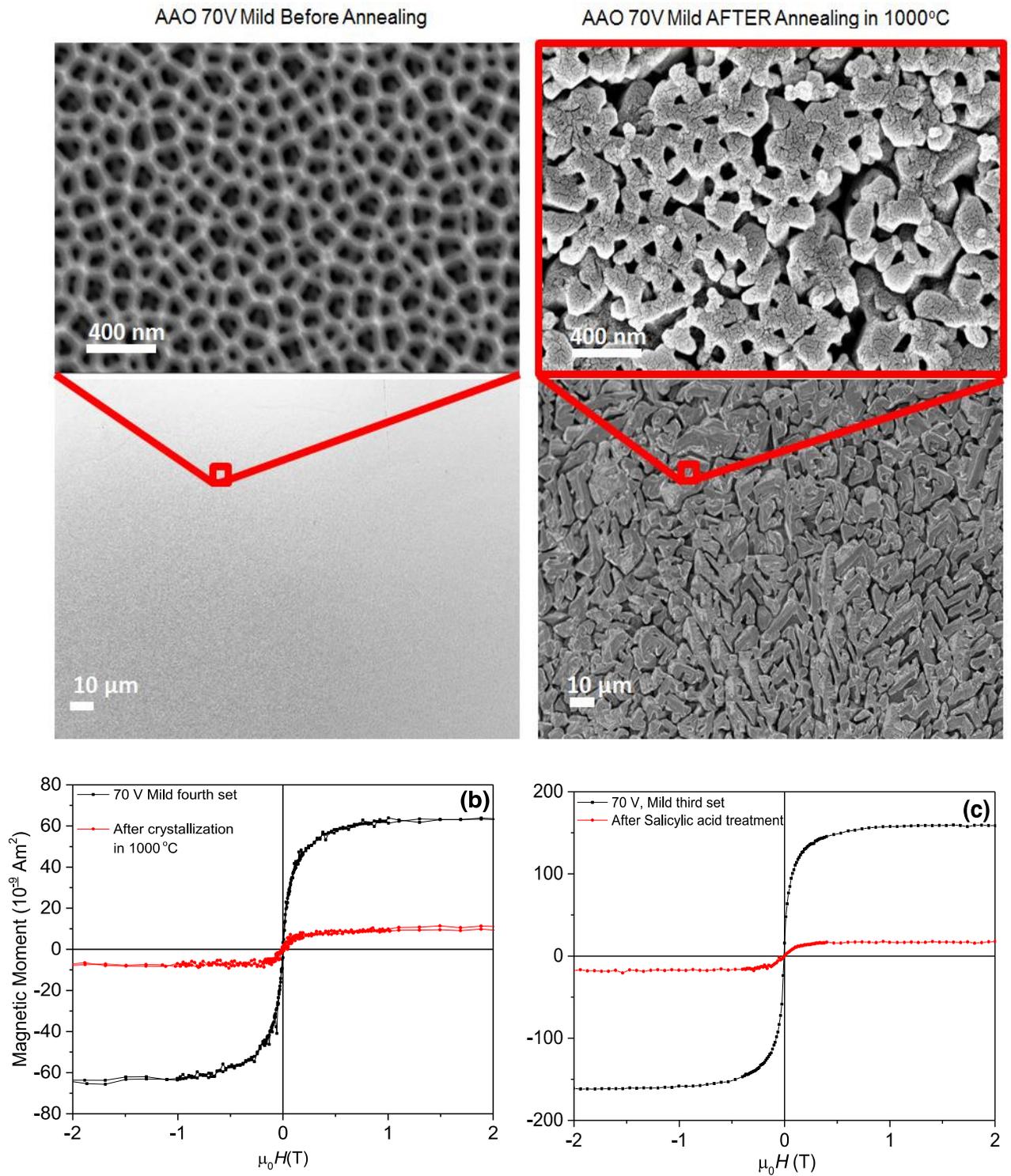

Figure 8. Structure (top) of a 70 V mild-anodized membrane before and after crystallization at 1000 °C. The panel at the bottom left shows the magnetization reduction on crystallization; the panel at the bottom right shows a similar drastic reduction of the magnetization produced by treatment of the membrane in salicylic acid.

The magnetic moment obtained from an AAO sample, is reduced by order of 30% after annealing in argon at 500°C for 6 hours. The Argon annealing intensifies the PL spectrum by factor of 3 [35].

Finally, figure 9 shows the UV-Vis spectra (transmission and absorption) for the 3rd set of AAO samples. New optical absorption appears and the bandgap decreases from about 5.6 eV to 3.5 eV on increasing the anodization voltage from the mild to the hard regime.

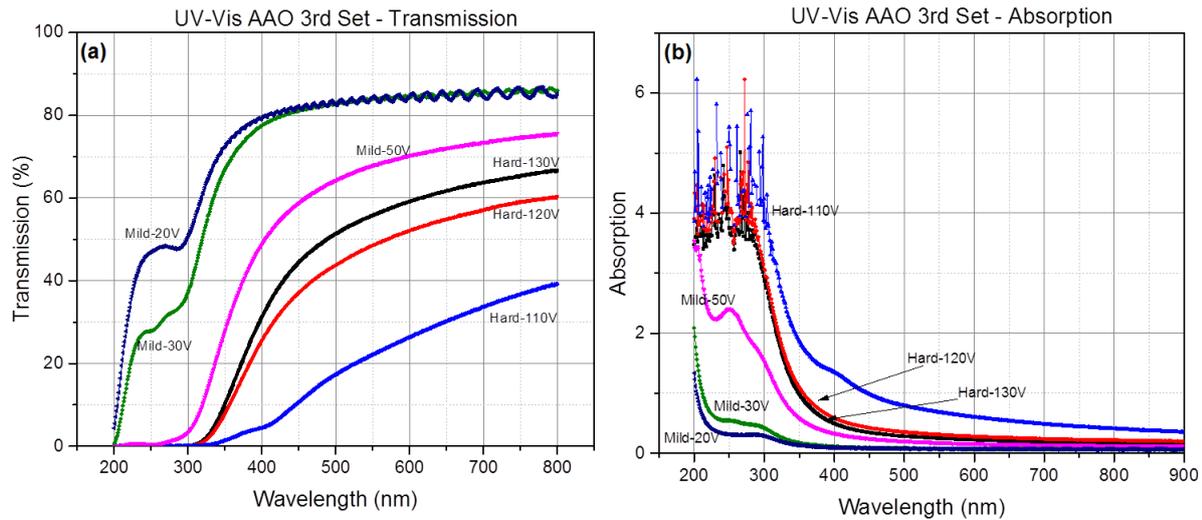

Figure 9 (a) Transmission and (b) absorption of a set of AAO samples

Table 1, Characteristics of the 3rd set of AAOs

| Anodization voltage | Type | Thickness | Magnetic Moment (Am$^2$) | Pore Size | Interpore Distance | ρ |
|---|---|---|---|---|---|---|
| 20 V | Mild | 3.32 μm | 6.2×10$^{-9}$ | 30 nm | 55 nm | 59.7 |
| 30 V | Mild | 7.7 μm | 7.3×10$^{-9}$ | 41 nm | 80 nm | 89.4 |
| 50 V | Mild | 27.3 μm | 1.7×10$^{-8}$ | 69 nm | 120 nm | 237.2 |
| 110V | Hard | 31.4 μm | 3.8×10$^{-8}$ | 57 nm | 260 nm | 48.0 |
| 120V | Hard | 35.4 μm | 4.2×10$^{-8}$ | 60 nm | 285 nm | 47.0 |
| 130V | Hard | 40.1 μm | 4.4×10$^{-8}$ | 62 nm | 310 nm | 46.9 |

## 4  Discussion

The data on these nanoporous structures establish the reality of *d*-zero magnetism, and allow us to characterize it better than was possible previously. The first question is 'Where in these membranes does the magnetism reside?' Since there is no such effect in well-crystallized alumina (sapphire) [36], and crystallization of the membranes tends to destroy the magnetic response (Fig. 8b), an origin connected with defects is clear. Anodic alumina has a highly-defective amorphous structure with an internal free volume of 10 – 20 %, in addition to the volume of the open nanopores. This is inferred by comparing the density extrapolated to zero open porosity (2945 kgm$^{-3}$ for mild anodization or 2570 kg m$^{-3}$ for hard anodization in oxalic acid [37]) with that obtained on model amorphous structures (3300 kgm$^{-3}$) [31].

A defect origin has been adduced for similar magnetic responses in oxide single crystals [8, 11], powders [15], thin films [38] and granular ceramics [5, 39-41] both in the undoped and doped states [2]. The random coordination structure of amorphous alumina that includes 4- 5- and 6-coordinated Al, and a mixture of $O^{2-}$ and $OH^-$ ions is a potential store of dangling bonds. However, the ability of solvents and drying treatments to modify the magnetism quite significantly was the clue that we should be looking for surface defects at the open pores rather than internal defects in closed pores. The salicylic acid treatment, which adds coordinating ligands to surface aluminium ions [34] is a convincing confirmation that the origin of the magnetism must lie at the surfaces of the open pores in the membranes.

Next, we ask what is the magnitude of the moment per unit pore surface area? The largest moment, observed in a 25 mm$^2$ 140 V hard-anodized membrane with $t$ = 41 µm, $d$ = 330 nm and $r$ = 30 nm, is 1.5×10$^{-7}$ Am$^2$ (Fig 7). From (1) we deduce $\rho$ = 41, and a moment density of 8 $\mu_B$ nm$^{-2}$. The largest moment in the mild anodized membranes is 1.1×10$^{-7}$ Am$^2$ for a 70 V membrane with $t$ = 71 µm, $d$ = 175 nm and $r$ = 50 nm, corresponding to $\rho$ = 398, and a moment 0.6 $\mu_B$ nm$^{-2}$. The surface moment densities are much greater for the more rapid, hard anodisation process (Fig 7b). Moment densities are similar to those found for the anhysteretic, temperature-independent surface magnetism of SrTiO$_3$ [11], due to localized or itinerant electrons associated with defect states at the pore surfaces. Oxygen vacancies are likely defects. It is widely believed that F-centres (oxygen defects with two trapped electrons) and F$^+$ centres (oxygen defects with one trapped electron) are responsible for the photoluminescence spectra [35, 42-49], and the EPR spectra have been attributed to F$^+$ centres [43, 49] since the two electrons in the F centre are expected to be spin paired. The Table summarises the unpaired spin densities deduced from our EPR measurements, compared with the unpaired spin densities that would be needed to explain the saturating moments. The number of unpaired spins in EPR, or those deduced from the Curie-law upturn in susceptibility and those needed to account for the saturating moments are quite different in magnitude, and indeed there is no reason to expect them to be related. If there is an ordered spin magnetization in the membranes, it will not give a sharp EPR signal, but rather a broad ferromagnetic resonance. Nor would it be expected to give a PL signal. A demonstration that the PL response and the *d*-zero magnetism are essentially unrelated was provided by heating the membranes to 500°C in argon. This is known to enhance the PL signal by a factor of 3 [35], but there was no corresponding increase in magnetic moment; it decreased by 30%.  We are dealing with two separate and independent electronic systems.

Table 2. Unpaired spin densities deduced from EPR and magnetization measurements.

| Voltage | spins/g | density (ppm) | $m$ (Am$^2$) | moment density (ppm) |
| --- | --- | --- | --- | --- |
| 20 | $2.4 \times 10^{16}$ | 2.0 | $1.5 \times 10^{-9}$ | 89 |
| 70 | $4.6 \times 10^{17}$ | 39.0 | $70 \times 10^{-9}$ | 149 |
| 140 | $7.4 \times 10^{17}$ | 62.6 | $80 \times 10^{-9}$ | 307 |

The nonlinear, saturating magnetic signals associated with *d*-zero oxide surfaces or thin films are frequently assumed to be evidence of ferromagnetism, and therefore potentially interesting for spintronics [1]. This is probably an unwarranted assumption. It is generally impossible in these systems to establish a reversible ferromagnetic-paramagnetic Curie temperature. To explain the observed temperature-independence of the magnetization curves in terms of spin-based ferromagnetism would require a Curie temperature at the pore surface greater than about 1000 K. To the best of our knowledge, there is no example in the literature of any material with itinerant or localized electrons having s = ½ that exhibits ferromagnetism at even a tenth of this temperature.

The absence of magnetocrystalline anisotropy is understandable in an s = ½ system, but shape anisotropy would be expected if ferromagnetism were confined to a thin surface volume.  For example, the maximum observed moment of 8 $\mu_B$ nm$^{-2}$ can be formally associated with a surface current around a pore of radius *r* = 30 nm, corresponding to an average magnetization of $(2/r)\sigma_m$ = 4.9 kAm$^{-1}$, where $\sigma_m$ is the surface moment density in A. It follows that the saturation of the magnetization in a field perpendicular to the plane should saturate practically immediately, but a field of 2.5 kAm$^{-1}$ (3 mT) is needed in an in-plane direction, where the demagnetizing factor is $\mathcal{N}$ = ½. Dipolar interactions between neighbouring cylinders will reduce the predicted anisotropy. Such a small difference in slope of the initial magnetization curves could not be detected in our measurements.

It is important to emphasize that the membranes are not superparamagnetic. While superparamagnets exhibit no hysteresis and saturate in relatively low magnetic fields, their magnetization curves follow a Langevin function, and scale as a function of ($H/T$) [33]. The initial slope in an external field should increase as $1/T$ until it is limited by the demagnetizing field. In the alumina membranes there is no temperature dependence at all, and no sign of blocking down to 4 K. Magnetically-ordered clusters of electron spins localized at oxygen vacancies, for example, would have expected to behave superparamagnetically.

A further indication that the magnetic response may not be associated with a collectively ordered magnetic state of the spins of electrons associated with surface defects on the pore surfaces comes from the EPR spectra. The spin density estimated from the intensity of the narrow EPR line (width 1.5 mT) measured on a 70 V membrane is $1.1 \times 10^{24}$ m$^{-3}$. These electrons are paramagnetic, not ferromagnetically ordered.

An alternative possibility is that source of the magnetism is quite unrelated to any collective ferromagnetic ordering of exchange-coupled spin moments. The idea is that we are seeing a new type of paramagnetism that is entirely field-induced and originates in coherent charge currents in mesoscopic electronic domains that form in response to the zero-point fluctuations of the vacuum electromagnetic field. It was shown in [50] that such a stable coherent state was possible in theory provided the system is quasi two-dimensional, with a large surface-to-volume ratio. Treated as a two-level system, the zero-point field of frequency ω mixes ground and excited electronic states to create a coherent many-electron ground state which is stabilized by an energy $-G^2\hbar\omega$ per electron, where G ~ 0.1. The ground-state wavefunction of the many-electron coherent state does not have a well-defined angular momentum, as it is based on the sum of two states with different energy and angular momentum values [19]. In the present case, the range of ω can be determined by cavity modes in the pores of length $t_m$ where the relevant electrons are located. Allowed modes have frequency $\omega_n = 2\pi nc/t_m$, with a low-frequency cutoff at a wavelength λ = $t_m$. A range of chemical effects of the zero-point fluctuations on systems confined in various cavities have been demonstrated recently by Ebbesen and co-workers [51-54].

Orbital paramagnetism has been invoked in the context of Au nanoparticles [16-18]. The effect of an applied magnetic field in our case is to modify the coherent state and induce giant orbital paramagnetism of electrons in the coherent domains. We have extended the theory based on coherent domains of non-interacting spinless electrons [50] to explain the anhysteretic, temperature-independent saturating paramagnetic response of $CeO_2$ nanoparticles. The mixing in the coherent state is modified by the static magnetic field, and the magnetic response is obtained by evaluating the expectation value of $\Sigma\mu_c\cdot\mathbf{B}$, where $\mu_c$ is the magnetic moment per coherent electron. Details can be found in the supplemental information to [19], where magnetization curves are fitted to the theoretical expression

$$M = M_s\, x/\sqrt{(1 + x^2)} \qquad (3)$$

where $x = C\,\mu_0 H$ Fitting the magnetization curves of the alumina membranes yields values of $C = 6.0 \pm 1.1$ T$^{-1}$ for the mild samples and slightly larger values $6.7 \pm 1.2$ T$^{-1}$ for the hard ones. The characteristic wavelength $\lambda$ of the radiation involved is related to $C$ in the theory by

$$\lambda = [(C/M_s)(6\hbar c\, f_c)]^{1/4} \qquad (4)$$

where $f_c$ is taken as the volume fraction of the sample that is magnetically coherent. An upper estimate of $f_c$ in these membranes is the alumina volume fraction $f_c = 1 - 2\pi r^2/3d^2$. For the most-magnetic mild-anodized membranes, $M_s \approx 60$ Am$^{-1}$ and $f_c \approx 0.83$; Equation 4 yields $\lambda = 338$ nm. For the most-magnetic hard-anodized membranes, $M_s \approx 100$ Am$^{-1}$ and $f_c \approx 0.98$; $\lambda = 338$ nm. It is very interesting that a comparison of the UV absorption spectra of strongly and weakly magnetic and nonmagnetic membranes exhibit absorption in the 200-300 nm range (Fig. 9). These wavelengths may be resonant with cavity modes in the pores, provided $t_m > \lambda$. The order of magnitude of the number of electrons $N$ in a coherent domain is $2\pi r\lambda\sigma_e$ where $\sigma_e$ is the electron density associated with oxygen defects at the pore surface. There are 18 oxygen ions per square nanometer, so a value of 1 – 10 electrons per square nanometer for $\sigma_e$ would give $N \sim 10^5 - 10^6$.

Finally, we consider the effect of changing the direction of magnetic field. The observation is that there is no difference in the magnetic response regardless of whether the field is applied in-plane or perpendicular to the plane. We could not reproduce the results of Sun et. al. [25] for very brief mild anodisation at 40 V in oxalic acid. The theory of the magnetism [19] is based on quantum-mechanical mixing of the ground and excited states in the applied magnetic field, neither of which corresponds to a well-defined orbital state of the system. It follows that there is no preferred direction and no anisotropy of the paramagnetic magnetization process beyond that due to shape anisotropy, which we have seen should be negligible on account of the very small magnetization.

In summary, the model of giant orbital paramagnetism of electrons in extended coherent domains is a simplification, but it gives a fair and reasonable account of the physics of the phenomenon of *d*-zero magnetism in these membranes.

## 5. Conclusions

Our experimental study establishes the existence of *d*-zero magnetism in well-characterized nanoporous anodic alumina membranes. The magnetism is anhysteretic and temperature independent. It arises from the open pore structure, and it is due to electrons associated with oxygen defects at the internal pore surfaces. The effect is largely destroyed by crystallization of the membranes or by a chemical treatment that modifies the surface structure of the amorphous alumina.

No model of conventional collective spin-based ferromagnetic order seems able to account for the results. In most of the magnetic samples, the saturation moments are ≤ 1 $\mu_B$ nm$^{-2}$, which is much too low a spin density to allow exchange coupling of the strength required for a Curie temperature that could explain the temperature independence of the magnetization below room temperature. The new model of saturating giant orbital paramagnetism of electrons in coherent mesoscopic domains is able to account for the magnetization curves, and it predicts optical absorption in the UV around 200 - 300 nm, in accord with observations. Magnetism in this spinless model is not spontaneous; it is induced by the applied field, and results from a field-induced quantum mechanical mixing of the collective ground and excited states of the system, hence there is no temperature-dependence [19]. Spin-orbit coupling at the curved surfaces of the pores may well play a role, and it should be considered in future.

Acknowledgements. This work was supported by Science Foundation Ireland on grant 13/ERC/I256. ASE acknowledges a postgraduate fellowship from TCD. We are grateful to Prof. Plamen Stamenov, Prof. Balz Kamber and Dr. Asra Sadat Razavian for helpful discussions.